\documentclass[preprint,12pt]{elsarticle}

\usepackage{amssymb}
\usepackage{color,soul}
\journal{Nuclear Physics B}
\begin{document}

\begin{frontmatter}



\title{Phase-dependent heat current of granular Josephson junction for different geometries}


\author[label1]{Elaheh Afsaneh \corref{cor1}}
\ead{el.afsaneh@gmail.com}
%\author{Elaheh Afsaneh \corref{cor1} \fnref{label1}}
%\author{Elaheh Afsaneh\corref{cor1}}
%\author{Heshmatolah Yavari \fnref{label1}}
\author[label1]{Heshmatolah Yavari \corref{cor1}}
\address[label1]{Department of Physics, University of Isfahan, Hezar Jarib, Isfahan 81746, Iran}
%\tnotetext[label1]{Department of Physics, University of Isfahan, Hezar Jarib, Isfahan 81746, Iran}
\cortext[cor1]{Corresponding author. Tel.: +98 9171912101; fax: +98 3117922409}
%\cortext[cor2]{Corresponding author. Tel.: +98 9171912101; fax: +98 3117922409}

\begin{abstract}
We theoretically investigate the phase-dependent heat transport of a temperature-biased granular Josephson junction in the presence of a perpendicular magnetic field. We illustrate the influence of geometry of the junction on the thermal current. The use of granular Josephson junction rather than bulk one makes significant changes in the heat current behavior. The heat current diffraction pattern of the rectangular, circular and annular geometries with no trapped fluxons demonstrates similar to the current of s-wave superconducting junction. By increasing the number of trapped fluxon, the pattern of current behaves such as d-wave superconducting junction. The feasibility of using granular superconductors, with different geometries, controlled by the magnetic field provides an appropriate tool to obtain the desired result for a specific application.
\end{abstract}


\end{frontmatter}


%\section{}
%\label{}





%%%%%%%%%%%%%%%%%%%%%%%%%%%%%%%%%%%%%%%%%%%%%%%%%%%
\section{Introduction}
%%%%%%%%%%%%%%%%%%%%%%%%%%%%%%%%%%%%%%%%%%%%%%%%%%%
 Advances in condensed matter physics and technology have provided the significant progress on the thermal transport of nanosystems\cite{Giazotto-2006,Dubi}. The improvement of heat current has proposed the quantum heat machines\cite{Broeck}, quantum refrigerators\cite{Venturelli} and thermoelectronic devices\cite{Roberts}. In the past decades, the heat current through the Josephson junction has been attracted much interest. 
 
For the first time Maki and Griffin proposed the interference term in addition to the quasiparticle one for the heat current through the Josephson junction\cite{Maki}. It was predicted that the interference current was depending on the superconducting phase and was due to an interplay between the quasiparticles and cooper pairs. For years, a plenty of projects was proposed to demonstrate the anomalous interference term\cite{Guttman-97,Guttman-98,Zhao-2003,Zhao-2004}. In spite of extensive attempts, no experiment could observe this phase-dependent term until 2012. Ultimately, F. Giazotto and M. J. Martinez-Perez proved the predicted phase-dependent term of thermal current in a heat interferometer dc-SQUID experiment\cite{Giazotto-Nature}. 
The modulation of phase-dependent thermal current through the temperature-biased Josephson junction was analyzed by means of magnetic flux similar to the electrical current through voltage-biased Josephson junction\cite{Giazotto-2012}. 
%Deleted%%%%%%%%%%%%%%%%%%%%%%%%%%%%%%%%
%The tunable superconducting phase difference by using the external magnetic flux has been paid more attention to be employed as the control elements of circuits\cite{Saira,Pekola}.  
%Deleted%%%%%%%%%%%%%%%%%%%%%%%%%%%%%%%%
In a temperature-biased Josephson junction, thermal current diffraction patterns were observed in a flux driven junction for the first time\cite{Giazotto-2014}. 
%Added-2%%%%%%%%%%%%%%%%%%%%%%%%%%%%%%%% 
Manipulation of heat currents based on phase-coherent caloritronics devices was proposed for several nanostructures\cite{Martinez-2014}and was investigated by mastering the superconducting quantum phases in temperature-biased Josephson junctions\cite{Fornieri}. 
Recently, thermal hysteresis behaviors were discussed in temperature-biased SQUID to provide thermal memory devices\cite{Guarcello-2017, Guarcello-2018-1}. 
%Added-2%%%%%%%%%%%%%%%%%%%%%%%%%%%%%%%%
%Deleted-2%%%%%%%%%%%%%%%%%%%%%%%%%%%%%%%%
%Recently, the possibility of controlling thermal currents is provided by the implementation of nanoscience technology in phase-coherent caloritronics\cite{Martinez-2014, Fornieri}. The superconducting circuits with phase-coherent caloritronics composed of thermal devices such as thermal modulators, heat transistors, phase-coherent thermal splitters, heat engines and thermal valves have been interested extensively. The phase-coherent caloritronics is able to master the heat currents based on tuning the phase difference of the superconducting junctions\cite{Blanc}. Fast caloritronic devices provide manipulating local temperature and heat power in procedure of soliton control through the Josephson junction. In this strategy, heat oscillators are propose to be applied in nano-heat engines and coherent-heat machines\cite{Guarcello-arxiv}. In a temperature-biased SQUID containing a non-negligible ring inductance, the modulation of heat transport with the external magnetic flux shows hysteresis behavior\cite{Guarcello-2017}. Very recently due to the thermal hysteresis behavior of a temperature-biased SQUID under the flux control, a thermal memory device has been proposed\cite{Guarcello-2018-1}. This superconducting thermal memory can provide logic memory states which results from the coherent heat currents through the Josephson junction. 
%Deletted-2%%%%%%%%%%%%%%%%%%%%%%%%%%%%%%%%

In order to improve the transport properties through the Josephson junction, the granular superconductor can be applied rather than the bulk one. 
Recently, a great deal of interest has been paid to understanding the properties of the granular superconducting systems\cite{Granato, Jardim, Mallett}. 
Different characteristics of electron transport and other electric responses to the external field have been studied on the superconducting granular systems\cite{sergeenkov}.  

The two-dimensional granular superconductor was arranged in honeycomb structure to investigate the phase oscillations\cite{banerjee}. A granular multilayer of superconducting-ferromagnetic structure was supposed to achieve the proximity effect\cite{Greener}. The characteristics of a superconducting granular structure were demonstrated by fluctuation spectroscopy close to the critical temperature\cite{Klemencic}. In studying the transport properties of the d-wave granular superconducting system under the electric field, the critical current is increased by the applied strong electric field\cite{Dominguez}. 
A two-fluid model was proposed to describe the transport characteristics of the granular superconductors which was well agreed with the different high-Tc superconductors\cite{Santos}. 
For the weak coupling, the conductivity of granular superconductors was investigated in the insulating regime and it was found that the charging energy of each grain could grow up the superconducting gap magnitude\cite{Lopatin}. In two-dimensional granular superconductors, the Nernst effect was studied using simulations with Langevin and RSJ dynamics\cite{Andersson-2010}. 

In recent years, the different characteristics of thermal transport in granular superconductors have been found much interest.  
%Added-1%%%%%%%%%%%%%%%%%%%%%%%%%%
In a temperature-biased long Josephson junction, it was shown that the maximum phase-dependent heat current behaves similarly to the superconducting critical current\cite{Guarcello-2016}. 
The length and the damping of LJJ affect the behavior of the diffractions patterns. 
The lobes configuration of the thermal transport diffraction patterns is strongly related to solitons. In turn, the number of solitons depends on the both length of junction and the intensity of the external magnetic field. In a thermally-biased LJJ, the influence of solitonic dynamics and excitations on the phase-coherent heat transport through the junction was studied\cite{Guarcello-2018-2}. In this study, new coherent caloritronics devices were proposed which are based on the motion of solitons and can be controlled by the external magnetic field. The interplay between phase-coherent caloritronics and solitonic dynamics was explored to introduce fast caloritronic devices providing the control of local temperature and heat power in solitonic manipulation procedure\cite{Guarcello-arxiv}. In this strategy, heat oscillators were proposed to be applied in nano-heat engines and coherent-heat machines.
%Added-1%%%%%%%%%%%%%%%%%%%%%%%%%%

%Added-1%%%%%%%%%%%%%%%%%%%%%%%%%%
Here to progress the heat transport through the thermally biased Josephson junction, we consider electrodes made by granular superconductors.
%Added%%%%%%%%%%%%%%%%%%%%%%%%%%
To this end, the effect of geometry on the granular superconductors was considered for both regular and irregular structures. 
The dynamics properties of thermoelectric effects and heat transport for two-dimensional granular superconductors were studied numerically under the influence of magnetic field\cite{Andersson-2011}. 

Another protocol to enhance the transport properties becomes possible by applying different geometric frustration for the junction areas. The particular geometries used in junctions for studying the electrical transport\cite{Afsaneh-s,Afsaneh-d} and thermal current\cite{Giazotto-2013} are rectangular, circular, and annular. 

The aim of this paper is to study the phase-dependent heat current of the granular Josephson junction under the effect of a magnetic field control. Previously, we calculate the electric transport of granular s-wave\cite{Afsaneh-s} and d-wave\cite{Afsaneh-d} superconducting systems in an applied magnetic field. To obtain the thermal current of the granular Josephson junction in analogy with the electric current, firstly we consider the heat current of the bulk superconducting system with different geometries\cite{Giazotto-2013}. After that by applying the Meilikhov \'s method\cite{Meilikhov}, we derive the thermal current of granular superconductor in rectangular, circular and annular geometries. 
 
This paper is organized as follows: In Sec.(\ref{Model}), we describe a model to obtain the thermal transport of a granular Josephson junction under the perpendicular magnetic field. In Sec.(\ref{Geometry}), we calculate the heat current of Josephson tunnel junction with different geometries for the bulk superconducting contacts. In Sec.(\ref{Granular-Current}), we derive the thermal current through the granular Josephson junction for the rectangular, circular and annular geometries. In Sec.(\ref{Results}) to represent the results of this study, we compare the plots of granular heat current with bulk one for the various geometries. In Sec.(\ref{Conclusion}), we conclude the obtained results in the present research. 

%%%%%%%%%%%%%%%%%%%%%%%%%%%%%%%%%%%%%%%%%%%%%%%%%%%
\section{Model}\label{Model}
\begin{figure}[h]
\centering
\includegraphics[scale=0.35]{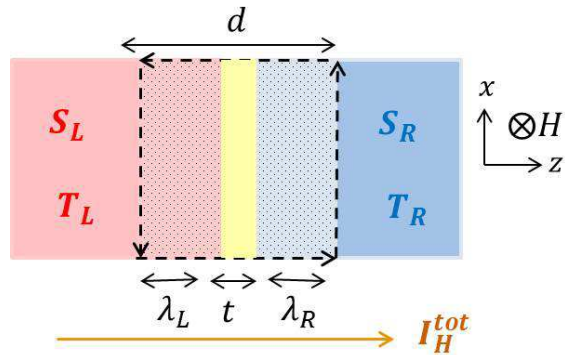}
\caption{ Josephson tunnel junction in the presence of perpendicular magnetic field. Pointed area shows the closed integration. $T_i$ and $\lambda_i$ indicate the temperature and London penetration depth of superconducting contacts $S_i$  ($i=L,R$). $t$ denotes the insulator thickness and $d=\lambda_L +\lambda_R +t $ represents the magnetic penetration depth.}\label{Fig1}
\end{figure}

%%%%%%%%%%%%%%%%%%%%%%%%%%%%%%%%%%%%%%%%%%%%%%%%%%%
The physical system under study is shown schematically in \ref{Fig1}. 
%%%%%%%%%%%%%
%%%%%%%%%%%%%
The proposed system is a long Josephson junction(LJJ) composed of a thin insulating barrier weakly coupled with two superconducting electrodes under the thermal bias. It means that the left and right leads are connected to the different heat baths with no bias voltage. Non-zero temperature difference between two contacts makes a heat current flowing through the junction. 
To neglect the effect of the edges, the Josephson junction is assumed symmetric. 
%Added-1%%%%%%%%%%%%%%%%%%%%%%%%%%%%%%%%
According to confinement of Josephson currents near the edges of junction, Josephson junctions can be identified into two classes of small and large ones. For LJJs, the edges of junction strongly confine the currents while in small junction, current distributes through the junction uniformly.    
A LJJ denotes a junction which has one dimension longer than the Josephson penetration length\cite{Barone,Tinkham}. The superconducting phase of this junction is a function of spatial coordinates. On the other hand, a short Josephson junction is a junction with dimensions smaller than the Josephson penetration depth which is assumed as point-like in space.
%Added-1%%%%%%%%%%%%%%%%%%%%%%%%%%%%%%%%

Usually, when a bias voltage is applied to the reservoirs with common heat bath, electrons transport from one lead to another which flows the electric current. 
The total electric current through the junction yields three contributios as follows: 
%%%%%%%%%%%
\begin{equation}\label{Itot}
I^{tot}(T_R,T_L,\varphi)=I^{qp}(T_R,T_L)+I^{int}(T_R,T_L) \cos(\varphi)+I^{Jos}(T_R,T_L)\sin(\varphi) 
\end{equation}
%%%%%%%%%%%%
where the first, second and third parts are respectively the quasiparticle, interference and Josephson terms of the electric current. Also, $\varphi=\varphi_L-\varphi_R$ denotes the phase difference of superconducting reservoirs. Here, we obtain the heat current in analogy with the electric one\cite{Langenberg,Eckern,Pedersen}. 

Particularly for the heat current, the superconducting condensate carries no entropy in static situation. In other words, the Josephson current term which represents the condensate Cooper pairs has no contribution in the heat transport\cite{Maki,Giazotto-Nature}. 
Therefore, when a temperature bias($ T_L>T_R$ ) is applied to the electrodes, a steady-state heat current containing two terms flows from the left side to the right(Fig.(\ref{Fig1})):
%%%%%%%%%%%%%%
\begin{equation}\label{I-H}
I^{tot}_H=I^{qp}_H (T_L,T_R)+I^{int}_H (T_L,T_R) \cos \varphi 
\end{equation}
%%%%%%%%%%%%%%
in which, $I^{qp}_H$ is the usual heat flux carried by quasiparticles\cite{Maki, Frank, Zhao-2004, Giazotto-2006} and $I_{int}(T_R,T_L) \cos(\varphi)$ denotes the interference term. The interference part of heat current as a function of the superconducting phase difference was predicted by Maki and Griffin\cite{Maki,Pedersen}.
 
The intrinsic superconducting phase-difference is influenced by the external magnetic field. So the only response of the heat current to the external magnetic field is the phase-dependent interference contribution.  The phase difference of system as a function of the applied magnetic field is specified by integration along the closed contour illustrated in Fig.(\ref{Fig1}) \cite{Giazotto-2014,Barone,Tinkham}:
%%%%%%%%%%%%%
\begin{equation}
\varphi(x)=\frac{\Phi}{\Phi_0}x+\varphi_0 
\end{equation}
%%%%%%%%%%%%%
where $\Phi$ denotes the magnetic flux, $\Phi_0$ indicates the quantum flux and $\varphi_0$ is the constant parameter. 
By integration over the junction area, the phase-dependent heat current is obtained: 
%%%%%%%%%%%%%%%
\begin{equation}
I_H(\Phi,T_L,T_R)=\int\int ds \rho_I(s,T_L,T_R)\cos(\varphi) 
\end{equation}
%%%%%%%%%%%%%%%
where  $\rho_I$ denotes the heat current density which is assumed constant. 
The definition of the heat current density varies for different geometries. 
\section{The geometry of Junction }\label{Geometry}
\begin{figure}[h] 
\centering
\includegraphics[scale=0.26]{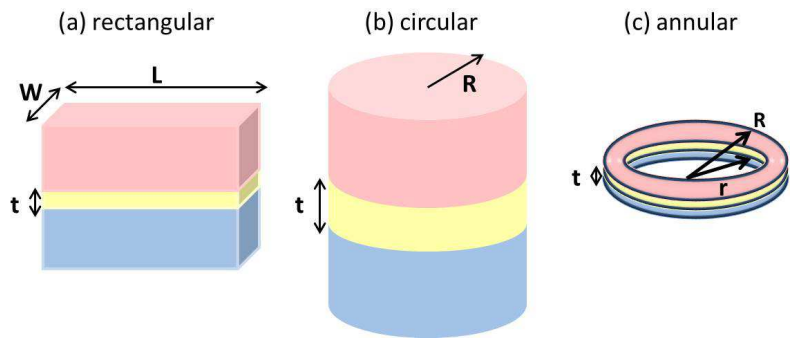}
\caption{Representative junctions: $(a) rectangular$, $(b) circular$, and $(c) annular$ geometries. $ L$, $ W$, $ R$, and $ r$ indicate the geometrical parameters of junctions and  $t$ denotes the junction tickness \cite{Giazotto-2013}   }. \label{Fig2}
\end{figure}

\begin{figure}[h]
\begin{center}$
\begin{array}{cc}
\includegraphics[width=53mm]{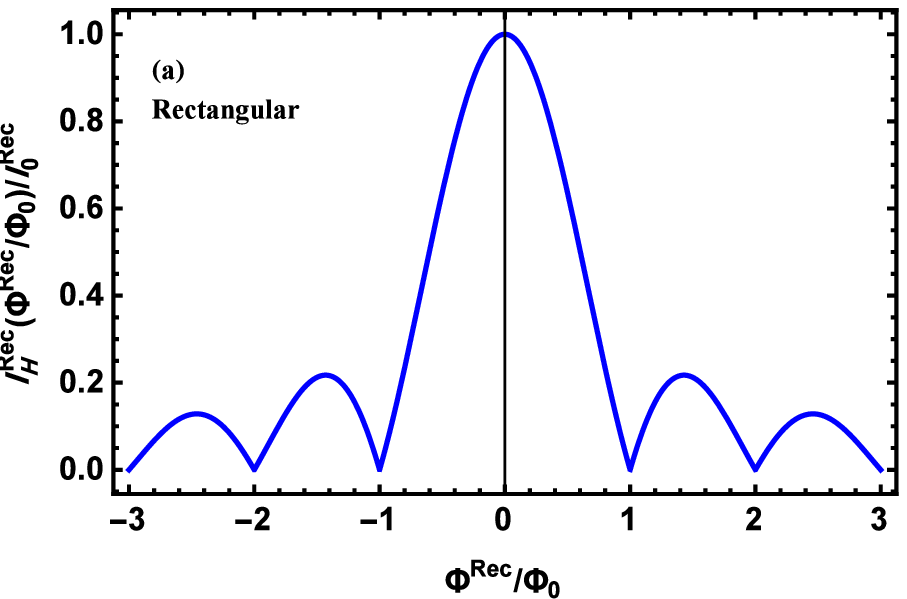}& \label{Fig3-a}
\includegraphics[width=53mm]{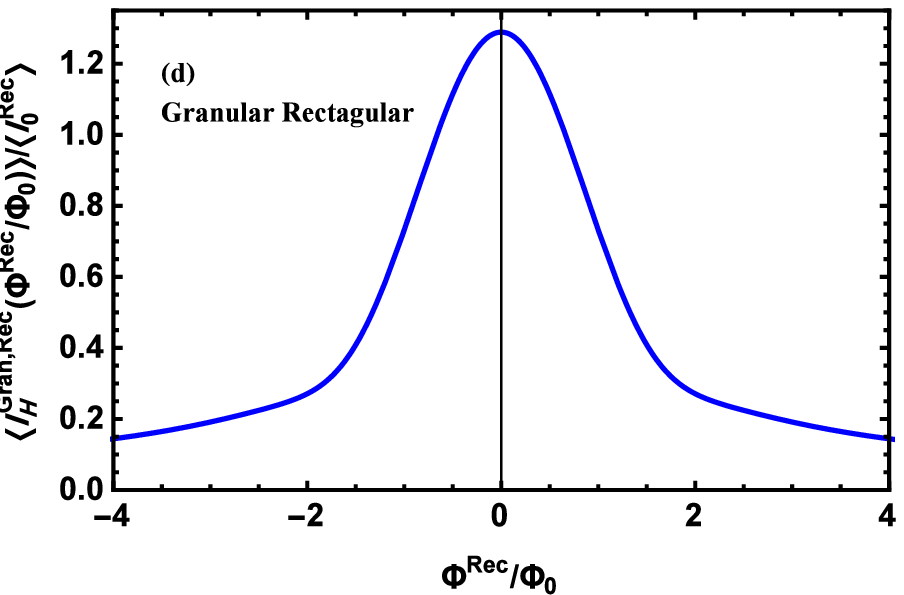} \label{Fig3-d}
\end{array}$
\end{center}

\begin{center}$
\begin{array}{cc}
\includegraphics[width=53mm]{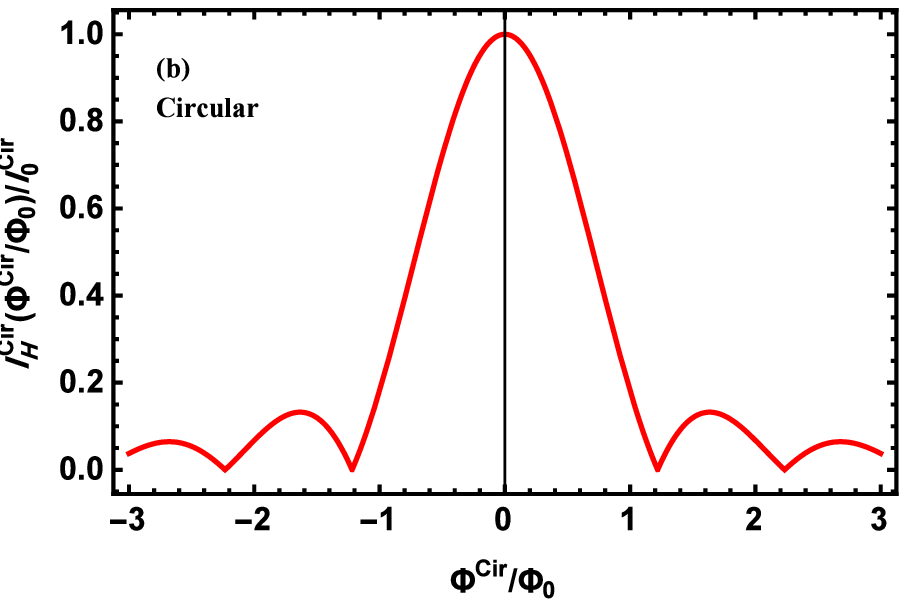}& \label{Fig3-b}
\includegraphics[width=53mm]{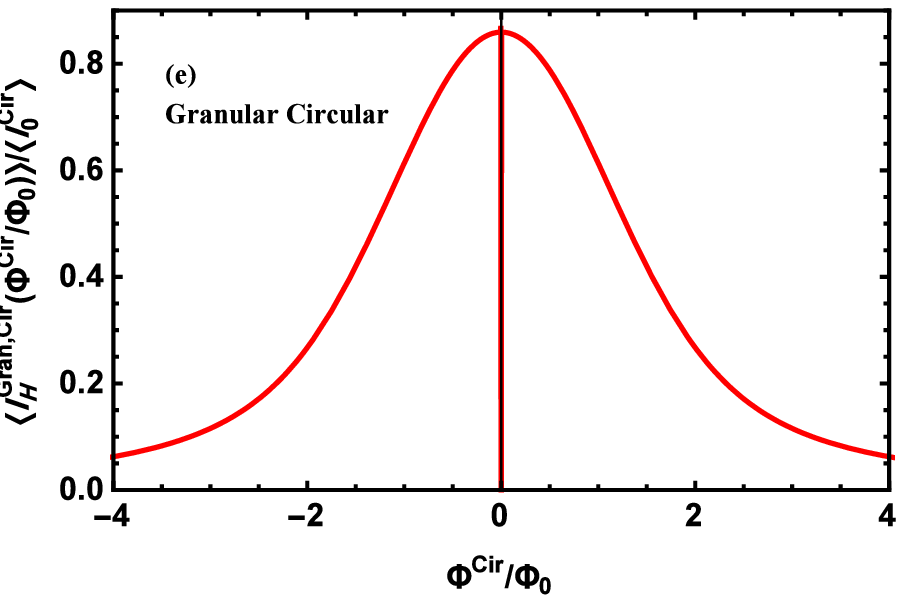} \label{Fig3-e}
\end{array}$
\end{center}

\begin{center}$
\begin{array}{cc}
 \includegraphics[width=53mm]{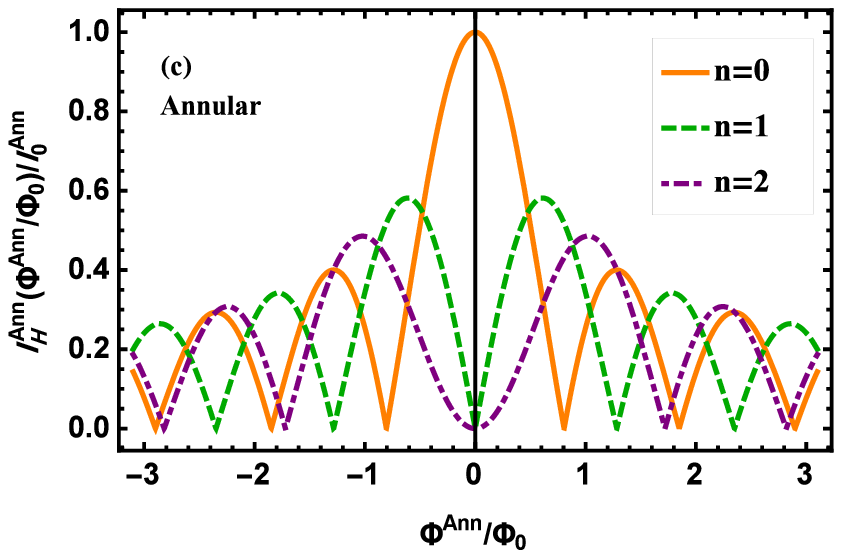}& \label{Fig3-c}
 \includegraphics[width=53mm]{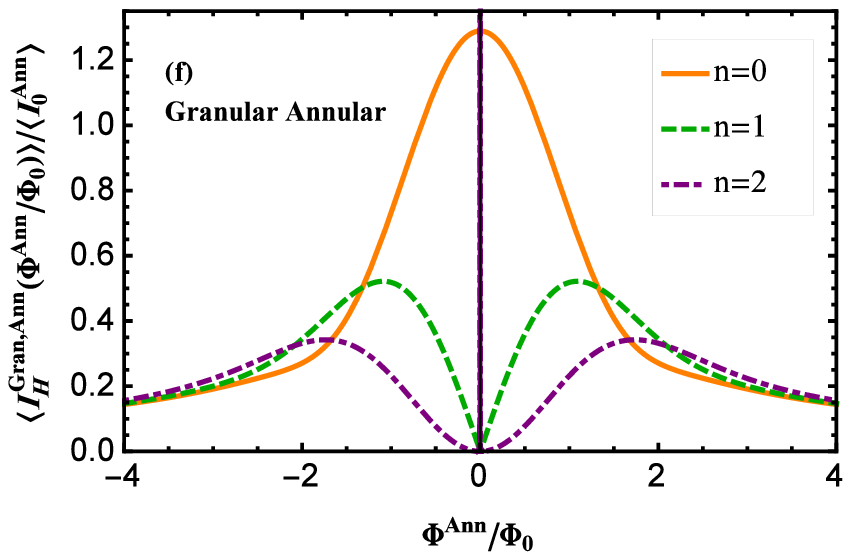} \label{Fig3-f}
\end{array}$
\end{center}
\caption{The phase-dependent heat current for left side: bulk superconducting electrods and right side: granular superconducting electrods junction in all geometries of (a) and (d): rectangular, (b) and (e): circular, (c) and (f): annular. For annular geometry, n indicates the number of fluxons trapped in the junction barrier and we set $\alpha= 0.9$.}
\end{figure}

%Removed%%%%%%%%%%%%%%%%%%%%%%%%%%%%%%%%%%%%%
%With respect to junction geometry,  junctions can be constructed in several types which are notably:  overlap, inline and annular geometries\cite{Pedersen-Soliton}. In the conventional linear LJJ such as rectangular and circular junctions, geometry of junction may disturb the reflections at the boundaries and the collisions of fluxons experimentally. So, the annular geometry is offered as a LJJ which is convenient to trap and store the magnetic flux of soliton quantization\cite{Davidson}.   
%Removed%%%%%%%%%%%%%%%%%%%%%%%%%%%%%%%%%%%%%
In the present study we investigate the heat current for three specific geometries, i.e., rectangular, circular and annular, sketched in Fig.(\ref{Fig2}). 
%%%%%%%%%%%%%
%%%%%%%%%%%%%%%%%%%%%%%%%%%%%%%%%%%%%%%%%%%%%%%%%%%%%
Here, we assume the uniform tunneling current distribution for all junctions. 
For rectangular geometry Fig.(\ref{Fig2})a, the thermal current respect to the magnetic flux is obtained\cite{Giazotto-2013}: 
%%%%%%%%%%%%%%%%%%%
\begin{equation}\label{Rectangular}
 I^{Rec}_H (\phi^{Rec}, T_L, T_R )=I^{Rec}_0 |\frac{\sin (\pi \frac{\phi^{Rec}}{\phi_0})}{\pi \frac{\phi^{Rec}}{\phi_0}}|
\end{equation}
%%%%%%%%%%%%%%%%%%%
where $I^{Rec}_{0}=\rho_I WL$ and $\phi^{Rec}=H_y L d$. Here, $W$ and $L$ are the dimention parameters of junction area and $d$ denotes the junction thickness.
%Added-1%%%%%%%%%%%%%%%%%%%%%%%%%%%%%%
The critical current which is influenced by the Josephson barrier characteristics can be modulated to provide the critical current diffraction patterns when an external driven field is applied. 
%Added-1%%%%%%%%%%%%%%%%%%%%%%%%%%%%%%
The pattern of heat current for the rectangular geometry is known as Fraunhofer diffraction which is illustrated in Fig.(\ref{Fig3-a})a.
%%%%%%%%%%%%%%%%%%%%%%
%%%%%%%%%%%%%%%%%%%%%%

The heat current of circular junction(Fig.(\ref{Fig2})b) is calculated as\cite{Giazotto-2013}:
%%%%%%%%%%%%%%%%%%%
\begin{equation}\label{Circular}
I^{Cir}_H(\Phi^{Cir}, T_L, T_R )=2 I^{Cir}_{0} |\frac{J_1(\pi \frac{\Phi^{Cir}}{\Phi_0})}{\pi \frac{\phi^{Cir}}{ \Phi_0}}| 
\end{equation}
%%%%%%%%%%%%%%%%%%%
where $\Phi^{Cir}=2H_y Rd$ and $I^{Cir}_{0}=\pi R^2 \rho_I $,  in which $R$ denotes the radius  of circle. Also,  $J_1(x)$  is the Bessel function of the first kind\cite{Arfken}. Fig.(\ref{Fig3-b})b demonstrates the heat current diffraction of circular geometry with the Airy pattern. 
%Added-1%%%%%%%%%%%%%%%%%%%%%%%%%%%%%%%
The ring-shaped annular LJJ offers great advantages in studying the soliton dynamics with no collision to the boundaries\cite{McLaughlin}. Soliton as a wavepacket can maintain its shape through the traveling with constant speed and colliding with others\cite{Rebbi,Doderer}. 

%Added-1%%%%%%%%%%%%%%%%%%%%%%%%%%%%%%%
It was recognized a long time ago that soliton or fluxon motion is smoother in the ring-shaped and also the fluxon is prevented to collide the boundaries, so the annular junction was suggested a unique chance to study the fluxon dynamics.
%Added-1%%%%%%%%%%%%%%%%%%%%%%%%%%%%%%%
In LJJs, the Josephson vortices can create magnetic flux quantum called fluxons or solitons which are able to move across the junction by an external bias.
%Added-1%%%%%%%%%%%%%%%%%%%%%%%%%%%%%%%
 Another unique property of the annular junction originates from the fluxoid quantization(fluxons). In otherwords, a superconducting ring as an annular junction acts like a fluxon trapping which can trap one or more fluxons inside the junction\cite{McLaughlin, Martucciello,Nappi}. 
 \\
Here, the heat current of the annular geometry is obtained(Fig.(\ref{Fig2})c)\cite{Giazotto-2013}: 
%%%%%%%%%%%%%%%%%%%
\begin{equation}\label{Annular}
I^{Ann}_H (\Phi^{Ann}, T_L, T_R )=\frac{2I^{Ann}_{0}}{1-\alpha ^2}|\int^{1} _{\alpha} dx x J_n(x\pi \frac{\Phi^{Ann}}{\Phi_0})| 
\end{equation}
%%%%%%%%%%%%%
where $I^{Ann}_0=\pi (R^2-r^2)\rho_I$ shows the heat current density and $\Phi^{Ann}=2H_y R d$ denotes the magnetic flux of annular geometry. In which $R$ and $r$ denote the external and  internal radius. Also, $J_n(x)$ defines the Bessel function of order $n$. The integer number of n indicates the trapped fluxons in the junction barrier. Moreover, $\alpha=\frac{r}{R}$ describes the lower limit of integrate. 

Particularly for the annular geometry, the diffraction pattern of the heat current drastically depends on the number of fluxons trapped inside the loop and is extensively different from the rectangular and circular ones which is illustrated in Fig.(\ref{Fig3-c})c. 
%Added-1%%%%%%%%%%%%%%%%%%%%%%%%%%%%%%%
The influence of the fluxons distribution on the critical current diffraction patterns of annular JJs has been studied experimentally\cite{Franz}. Moreover, a superconducting memory based on the hysteretical behavior of the Josephson critical current in LJJs, providing protection of the memory states against external perturbations, was suggested recently\cite{Guarcello-Sci-2017}.
%Added-1%%%%%%%%%%%%%%%%%%%%%%%%%%%%%%%

%%%%%%%%%%%%%%%%%%%%%
\section{Critical Heat Current of Granular Systems}\label{Granular-Current}
%%%%%%%%%%%%%%%%%%%%%%
In this section, we consider Josephson junctions formed by granular superconductors and calculate the granular heat current for this system. 
A granular superconductor consists of superconducting granule array while the grains weakly coupled with each other by electron tunneling an insulating substrate. The characteristics of superconducting array are determined by the properties of its grains\cite{Beloborodov}. According to the peculiarities of s-wave superconductivity, Anderson described a single small isolated grain\cite{Anderson}  which was completely agreed with experiments\cite{Ralph,Black}.

The conventional s-wave superconductors generally have no sensitivity to the disorderness. For these systems, the BCS theory is valid as the time-reversal invariance is not broken. Therefore, the critical temperature of a single grain has to be close to the bulk value which means that the critical temperature of granular system can be considered the same as the bulk magnitude\cite{Ovchinnikov-1,Ovchinnikov-2}. 
\\
%%%%%%%%%%%%%%%%%%%%%%%%%%%%%%%%%%%%%%%%%%
Based on the Meilikhov \'s model, the granular critical current is obtained by averaging over all superconducting grains with random distribution function\cite{Meilikhov}. Previously we implemented this procedure to obtain the electrical current of granular s-wave\cite{Afsaneh-s} and d-wave \cite{Afsaneh-d} superconductors with different geometries. 
Now in this study, we consider a Josephson junction made by conventional superconducting grains. Then, we calculate the granular heat current of this system in analogy with the granular electrical current under the method of Meilikhov\cite{Meilikhov}. To this end, we average over the random superconducting grains placed on an insulating matrix substrate with the Maxwell distribution. The granular heat current of the junction is obtained by integrating over the heat current of the single grain:
%%%%%%%%%%%%%%%
\\
\begin{equation}\label{Granular}
\langle I^{Gran,i}_H \rangle=\Big( \int \omega(r) (I^i_{Gran,i})^2 dr \Big)^{\frac{1}{2}}
\end{equation}
\\
%%%%%%%%%%%%%%%
here $i$ denotes the geometry of junction as: rectangular($Rec$), circular($Cir$) and annular($Ann$). The assumed distribution function, $\omega(r)$ is defined as:
%%%%%%%%%%%%%%%
\\
\begin{equation}
\omega(r)=\frac{32r^2}{\pi^2 a^3}e^{-\frac{4r^2}{\pi a^2}} 
\end{equation}
%%%%%%%%%%%%%%
\\
%%%%%%%%%%%%%%%%%%%%%%%%%%%%%%%%%%%%%%%%%%%
where $r$ denotes the radius of each grain and  $a$ is the average granule size\cite{Ignatjev1994}. 
%%%%%%%%%%%%%%%%%%
 For the rectangular geometry, the weak link between grains is formed in the region of the plane segments\cite{Ignatjev1998}. In other words, the cross section of coupled grains is assumed with rectangular geometry. By this definition the radius $r$ is proportional to the granule size\cite{Shmidt, Meilikhov}. 
To obtain the heat current of granular superconductors with rectangular geometry, we put relation(\ref{Rectangular}) in equation (\ref{Granular}). For the rectangular geometry, the granular heat current as a function of the magnetic flux is illustrated in Fig.(\ref{Fig3-d})d.

For circular geometry to achieve the granular heat current, we use the relation(\ref{Circular}) in the integral(\ref{Granular}) which is indicated in Fig.(\ref{Fig3-e})e.  

The granular heat current of annular geometry can be obtained by substituting the expression(\ref{Annular}) into the integral(\ref{Granular}). But, the granular heat current integrand for each trapped fluxon can not be solved simply. So, we use an appropriate approximation which was applied for small annular junction\cite{Martucciello,Nappi}. Accordingly, we take into account the heat current of relation(\ref{Annular}) approximated as the following: 
%%%%%%%%%%%%%%%
\\ 
\begin{equation}\label{Granular-Annular}
I^{Ann}_H(\Phi^{Ann},T_L,T_R)=I^{Ann}_{0} |J_n(kR)|
\end{equation}
\\
%%%%%%%%%%%%%%%
here, $ I^{Ann}_{0}$ denotes the heat current density with the similar definition used in relation(\ref{Annular}). Also, $J_n(x)$ is the nth order of Bessel function. Inserting (\ref{Granular-Annular}) into integral(\ref{Granular}) gives us the granular heat current for annular geometry which is shown in Fig.(\ref{Fig3-f})f. 
%%%%%%%%%%%%%%%%%%%%%
\section{Results} \label{Results}
%%%%%%%%%%%%%%%%%%%%%
To specify the effect of granular superconductors on the heat current, we compare the diffraction patterns of heat current for Josephson junctions made by the conventional bulk superconductors with the granular ones in specific geometries of rectangular, circular and annular. This comparison is shown in five panels of Fig.(4). In panel (a) of this figure, we note that the maximum value of the heat current for the rectangular geometry is higher considering a granular junction.
%%%%%%%%%%%%%%%%%%%%%%
% \end{flushright}
%%%%%%%%%%%%%%%%%%%%%%
    \begin{figure}[h]
%\begin{flushright}$
\begin{center}$
\begin{array}{cc}
\includegraphics[width=53.2mm]{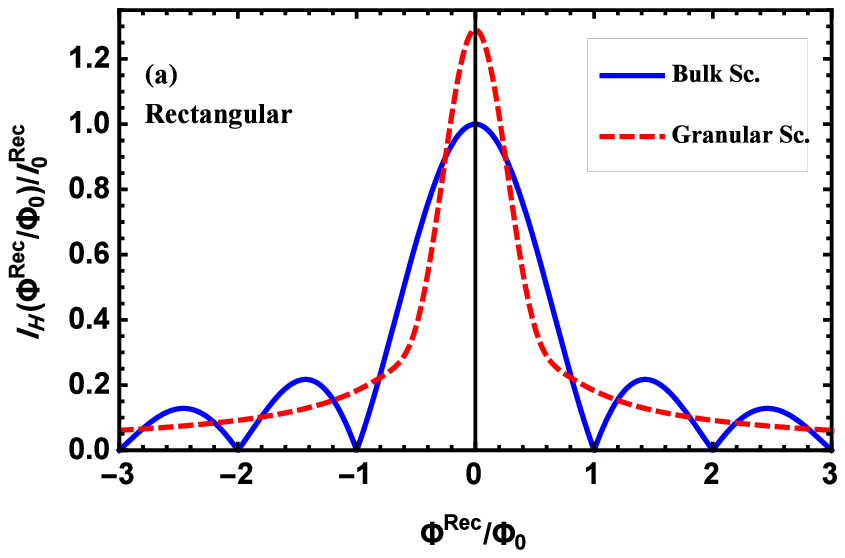}&
\includegraphics[width=53.2mm]{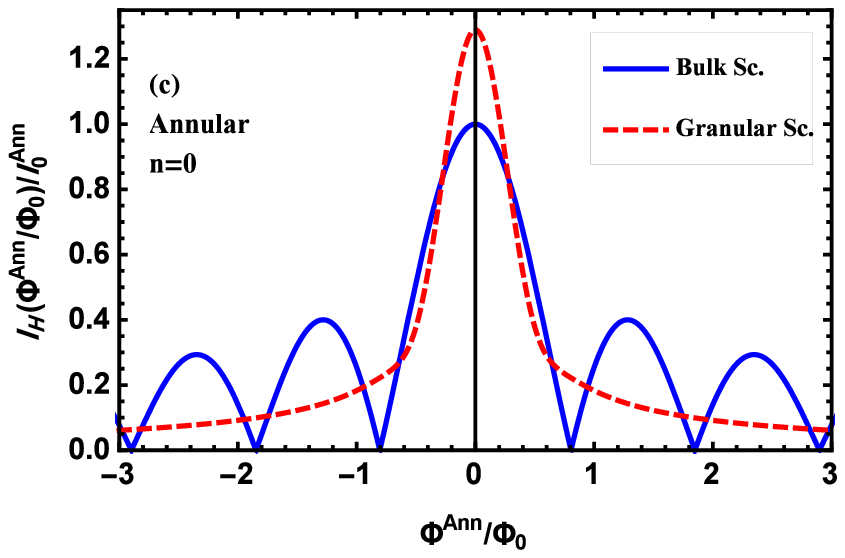}
\end{array}$
\end{center}
%\end{flushright}

%\begin{flushright}$
\begin{center}$
\begin{array}{cc}
\includegraphics[width=53.2mm]{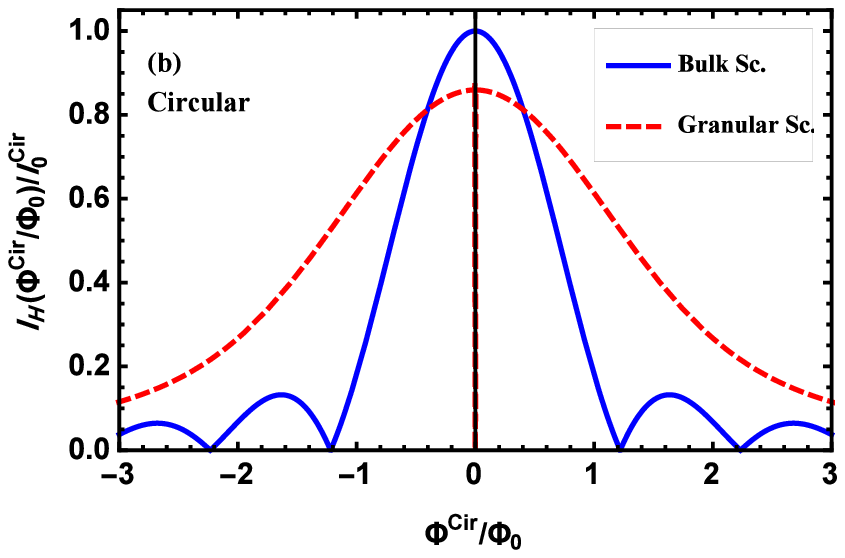}& 
\includegraphics[width=53.2mm]{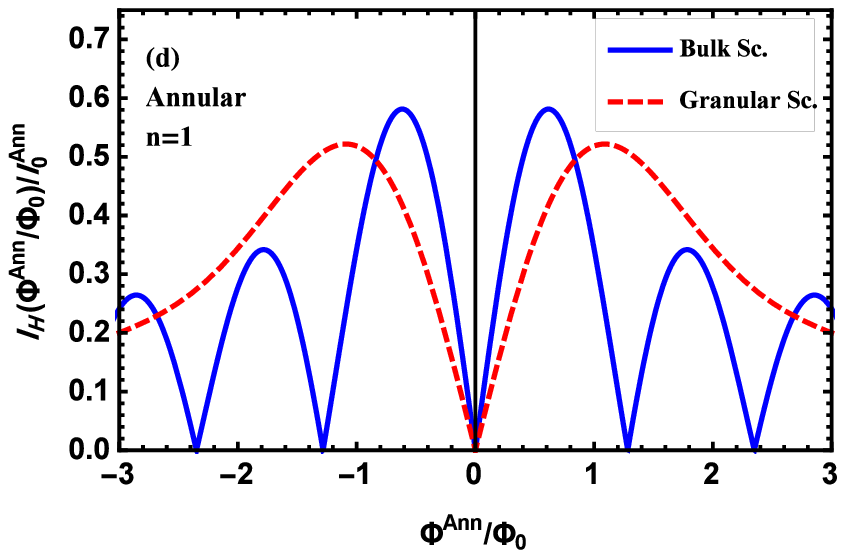}
\end{array}$
\end{center}
%\end{flushright}

%\begin{flushright}$
\begin{center}$
\begin{array}{rr}
&
 \includegraphics[width=53.2mm]{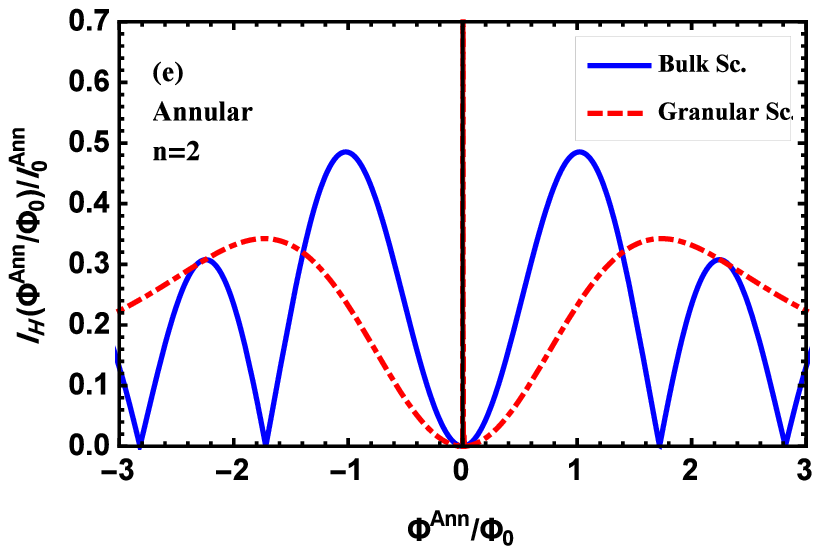}
\end{array}$
\end{center}
%\end{flushright}
\caption{Comparison between the heat current of bulk superconductors rather than granular ones for (a): rectangular geometry (b): circular geometry (c): annular geometry with n=0 (d): annular geometry with n=1 (e): annular geometry with n=2. }
\label{Fig4}
\end{figure}
 %%%%%%%%%%%%%%%%%%%%%%

For circular geometry although the heat current peak is higher for the bulk junction than the granular one, the latter covers much more area than the former which is indicated in Fig.(\ref{Fig4})b. 

In Fig.(\ref{Fig4})c, the heat current of the annular geometry with the lowest trapped fluxon reaches the higher point when the junction is granular. This behavior is similar to the rectangular and circular geometry unless it decays with faster rate. 
The plots of annular geometry(Fig.(\ref{Fig4})c-e) show that by increasing the number of trapped fluxon, the maximum point of the heat current for granular junction is reduced.         
Thoroughly, it is clear that the heat current of annular geometry with the lowest trapped fluxon(Fig.(\ref{Fig4})c) as well as rectangular(Fig.(\ref{Fig4})a) and circular(Fig.(\ref{Fig4})b) ones behave similar to the current of s-wave superconductors with only one peak. 
However for the higher number of trapped fluxon Fig.(\ref{Fig4})d and Fig.(\ref{Fig4})e, the heat current behaves such as the d-wave superconductor with two symmetric maximum peaks.   

%Added-2%%%%%%%%%%%%%%%%%%%%%%%%%%%%%%%%
Since the phase-dependent granular heat current of the proposed system depends on the superconducting phase difference, it can be controlled by the external magnetic field. Therefore, one of the most significant characteristics of this system is focused on the control ability of the thermal transport by the tunable magnetic flux. Also, the implement of the granular junction rather than the bulk one makes it possible to manipulate the peak height and the decay rate of the heat current. Furthermore, applying the various geometries for junction area can provide extensively changes on the heat current. Consequently by mastering the mentioned elements, the present system can be engineered to achieve the desired heat current relevant to its application.
%Added-2%%%%%%%%%%%%%%%%%%%%%%%%%%%%%%%%
 %%%%%%%%%%%%%%%%%%%%%%%%%%%%%%%%%%%%%%%%%%%%%%%%%%%
\section{Conclusion}\label{Conclusion}
%%%%%%%%%%%%%%%%%%%%%%%%%%%%%%%%%%%%%%%%%%%%%%%%%%%
In summary, we have studied the phase-dependent granular heat current as a function of a perpendicular magnetic field for different geometries. Particularly, we have compared the heat current diffraction patterns of bulk and granular junctions when the geometry of junctions are rectangular, circular or annular.  This comparison displays that we can achieve our desired heat current by engineering the system by means of the tunable magnetic flux, applying the granular superconductor and utilizing the various geometries.
%%%%%%%%%%%%%%%%%%%%%%%%%%%%%%%%%%%%%%%%%%%%%%%%%%%
\section{Acknowledgments}
The authors wish to thank the Office of Graduate Studies and Research Vice President of the University of Isfahan for their support.
%%%%%%%%%%%%%%%%%%%%%%%%%%%%%%%%%%%%%%%%%%%%%%%%%%%

\end{document}